\documentclass[prb,twocolumn,showpacs,preprintnumbers,amsmath,amssymb]{revtex4}
\usepackage{graphicx}
\usepackage{dcolumn}
\usepackage{bm}

\begin{document}

\title{Electronic structure of BaCu$_2$As$_2$ and SrCu$_2$As$_2$: sp-band metals}

\author{D.J. Singh}

\affiliation{Materials Science and Technology Division,
Oak Ridge National Laboratory, Oak Ridge, Tennessee 37831-6114}

\date{\today}

\begin{abstract}
The electronic structures of ThCr$_2$Si$_2$ structure BaCu$_2$As$_2$
and SrCu$_2$As$_2$ are investigated using density functional calculations.
The Cu $d$ orbitals are located at 3 eV and higher binding energy,
and are therefore chemically inert with little contribution near the
Fermi energy.
These materials are moderate density of states, sp-band metals with 
large Fermi surfaces and low anisotropy.
\end{abstract}


\maketitle

Recent results, motivated by the discovery of high temperature
Fe-based superconductivity, \cite{kamihara-a}
have shown that 3$d$ pnictides in the ThCr$_2$As$_2$ structure
show a remarkable range of properties.
This includes the itinerant spin density wave and high temperature
superconductivity characteristic of the Fe-based superconductors
for BaFe$_2$As$_2$, SrFe$_2$As$_2$ and CaFe$_2$As$_2$.
\cite{rotter-sdw,rotter-sc,sasmal,alireza,tegel,ronning}
Remarkably, in contrast to cuprates, superconductivity can be induced
by doping on the Fe-site, using Co and Ni as electron dopants.
\cite{sefat-co,matsuishi}
BaCo$_2$As$_2$ and BaNi$_2$As$_2$, show very different properties,
namely those of a material very close to ferromagnetism,
\cite{sefat-baco}
and a low temperature electron-phonon superconductor, respectively.
\cite{ronning-banias,kurita-banias,subedi-banias}

The electronic structures of BaFe$_2$As$_2$, BaCo$_2$As$_2$
and BaNi$_2$As$_2$ are, however, closely related,
\cite{subedi-banias,sefat-baco,nekrasov,singh-bfa}
with Fe $d$ bands near $E_F$, modest
hybridization with As, and a similar shaped density of states with
a pseudogap at an electron count of six $d$ electrons as in
the Fe-based oxyarsenides and FeSe. \cite{singh-du,subedi-fese}
The very different properties of these three compounds arise because of
the different electron counts of the transition elements, which lead to
different placements of the Fermi level. 
This provides an explanation for the ability to dope BaFe$_2$As$_2$
with Co or Ni and obtain a coherent superconducting alloy.
\cite{sefat-co}

The Mn and Cr compounds, BaMn$_2$As$_2$
and BaCr$_2$As$_2$, show rather different electronic structures
from BaFe$_2$As$_2$,
with strong spin dependent hybridization between the transition
element $d$ states and As $p$ states.
BaMn$_2$As$_2$ is an antiferromagnetic semiconductor,
\cite{an,ysingh}
while BaCr$_2$As$_2$ is an antiferromagnetic metal with itinerant character.
\cite{singh-cr}
It is interesting to note that although the electronic
structures and properties of the Fe compounds, FeSe and
BaFe$_2$As$_2$ are similar,
the trends with transition element substitution in the ThCr$_2$Si$_2$ structure
pnictides are very different from those in the $\alpha$-PbO
structure selenides, presumably reflecting the
different chemistry of pnictogens and chalcogens. \cite{ding-pbo}
Synthesis of BaCu$_2$As$_2$ and SrCu$_2$As$_2$ was reported
by Pfisterer and Nagorsen, \cite{pfisterer}
but little is known about their electronic structure or physical
properties.
Here we investigate the properties of the Cu compounds
BaCu$_2$As$_2$ and SrCu$_2$As$_2$ using density functional calculations.

The present first principles
calculations were done within the local density approximation
(LDA) using the general potential linearized augmented planewave (LAPW)
method, \cite{singh-book}
similar to prior calculations for BaFe$_2$As$_2$. \cite{singh-bfa}
We used the reported experimental lattice parameters, \cite{pfisterer}
$a$=4.446 \AA, $c$=10.007 \AA, for BaCu$_2$As$_2$ and
$a$=4.271 \AA, $c$=10.018 \AA, for SrCu$_2$As$_2$.
The internal parameter, $z_{\rm As}$ was determined by
energy minimization.
The resulting values were
$z_{\rm As}$=0.3657 for BaCu$_2$As$_2$ and
$z_{\rm As}$=0.3697 for SrCu$_2$As$_2$.
We used well converged basis sets, including local orbitals to
treat the semi-core states and relax the Cu $d$ state linearization.
\cite{singh-lo}
Relativistic effects were included at the scalar relativistic level.
The LAPW sphere radii were 2.2 $a_0$ for Ba and
Sr and 2.1 $a_0$ for Cu and As.

\begin{figure}[tbp]
\vspace{0.25cm}
\includegraphics[height=0.96\columnwidth,angle=270]{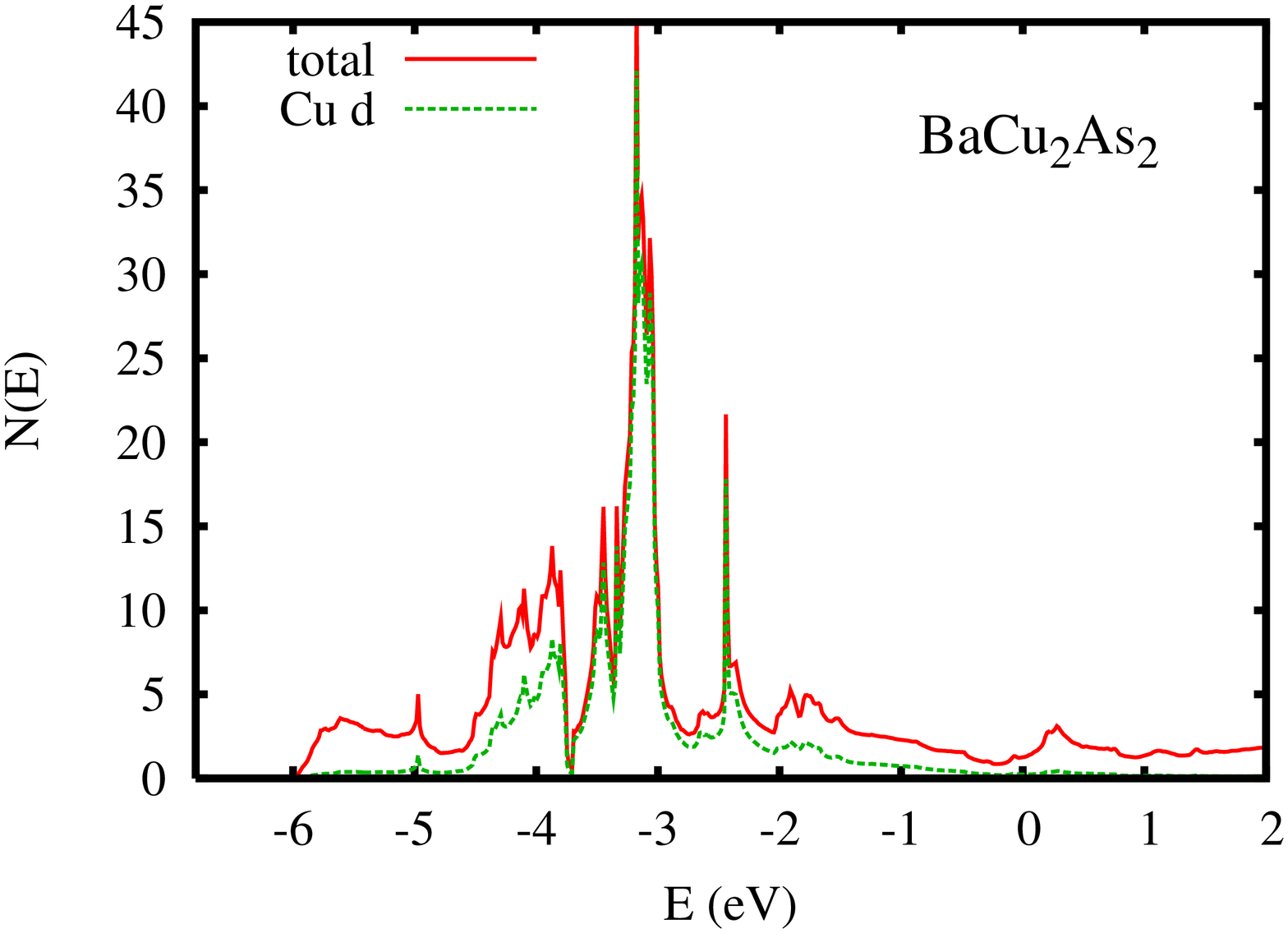}
\includegraphics[height=0.96\columnwidth,angle=270]{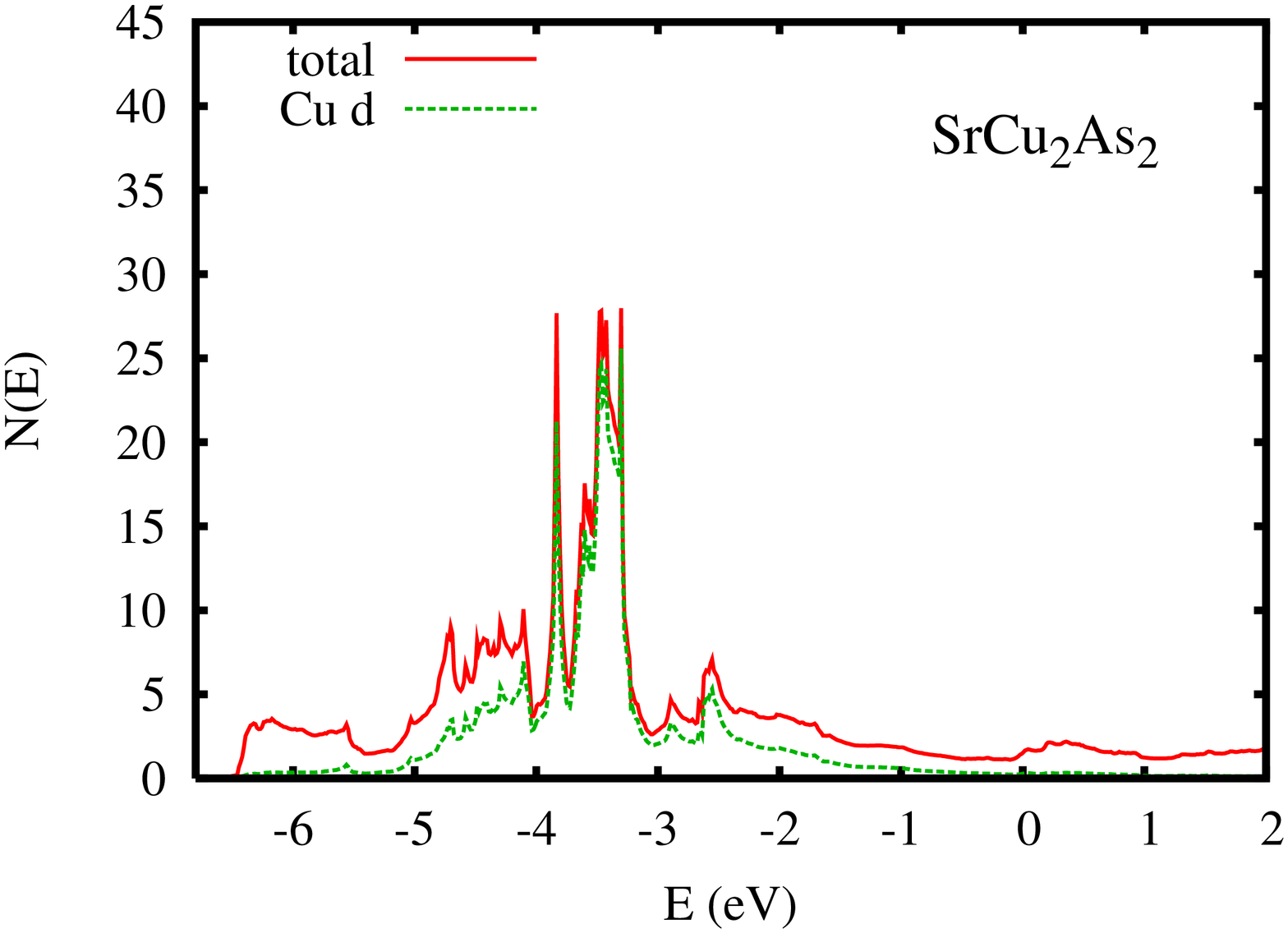}
\caption{(color online)
Electronic density of states and Cu $d$ projection
for BaCu$_2$As$_2$ (top) and SrCu$_2$As$_2$ (bottom).}
\label{dos}
\end{figure}

\begin{figure}[tbp]
\vspace{0.25cm}
\includegraphics[height=0.96\columnwidth,angle=270]{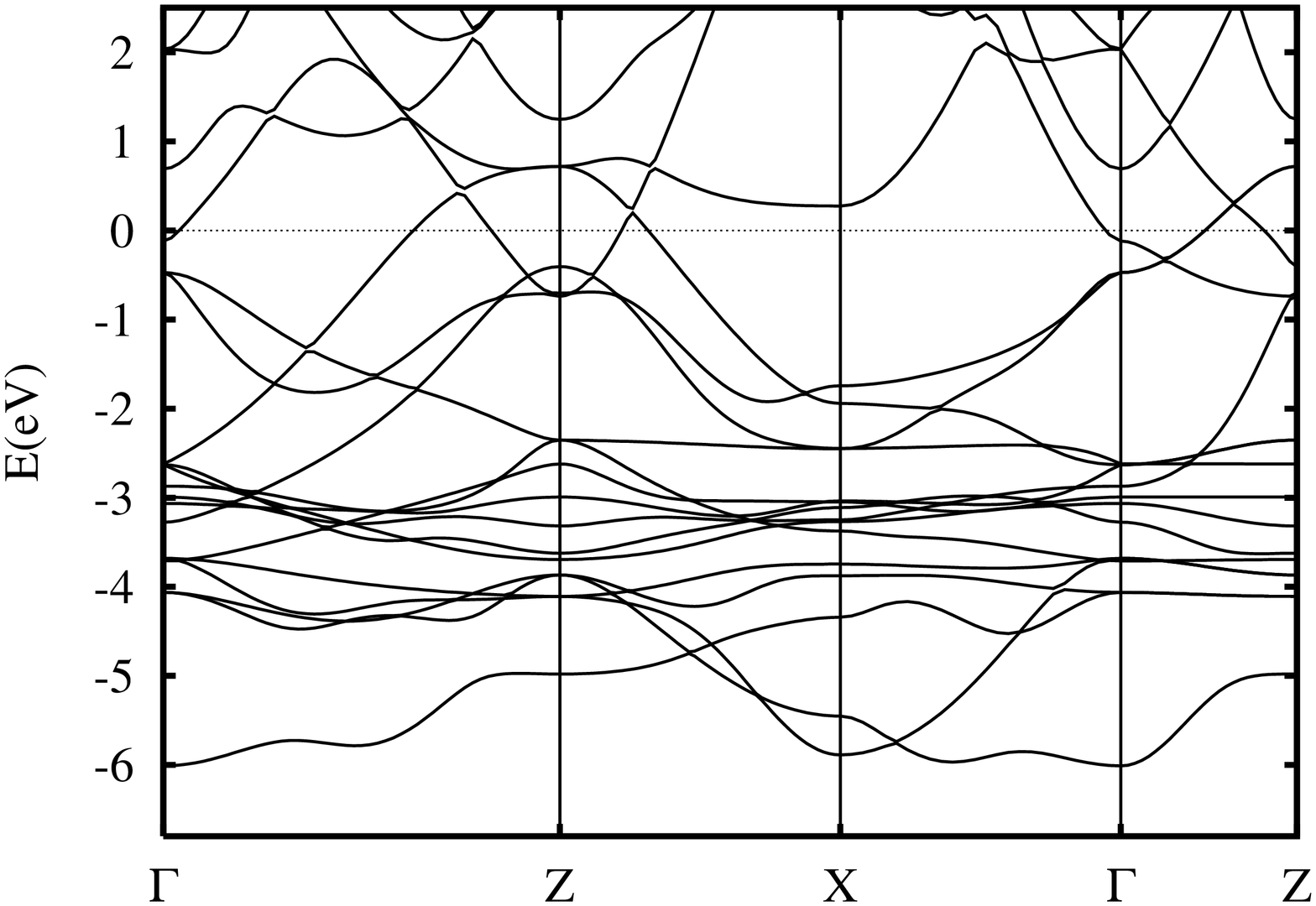}
\includegraphics[height=0.96\columnwidth,angle=270]{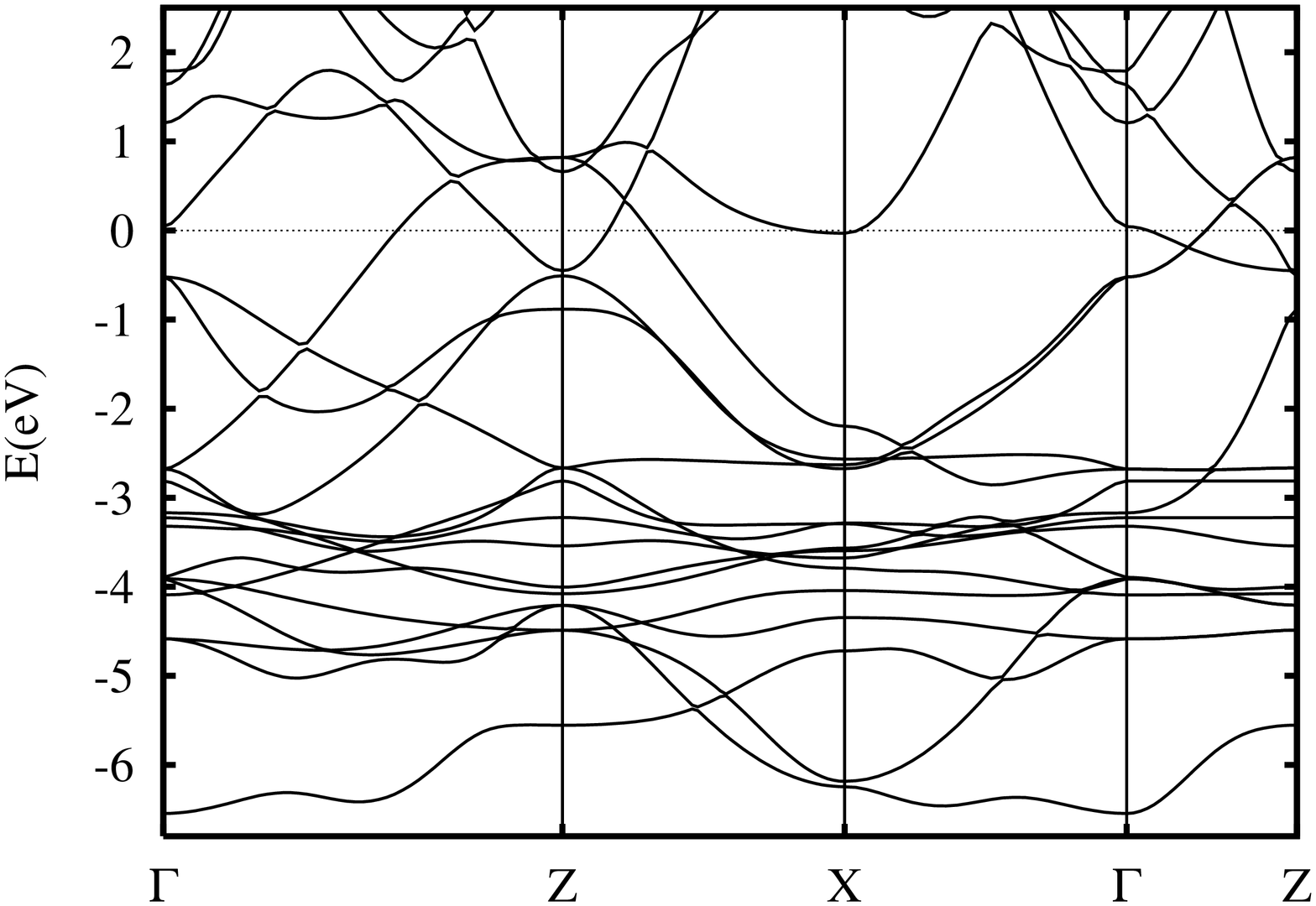}
\caption{
Band structure of BaCu$_2$As$_2$ (top) and SrCu$_2$As$_2$ (bottom).
The bands are plotted along lines in the basal plane and along the
$k_z$ direction.
The body centered tetragonal reciprocal lattice vectors are
($2\pi/a,0,-2\pi/c$),
($0,2\pi/a,-2\pi/c$), and ($0,0,4\pi/c$). In terms of these,
the long $\Gamma$-$Z$ direction is from 
(0,0,0) to (1,0,1/2) in the body centered tetragonal zone, while
the short $\Gamma$-$Z$ direction is from
(0,0,0) to (0,0,1/2). $X$ denotes the zone boundary (1/2,1/2,1/2) point.
A two dimensional band structure
would show no dispersion along the short $\Gamma$-$Z$
direction and would be symmetric about the mid-point of the long
$\Gamma$-$Z$ direction.}
\label{bands}
\end{figure}

The calculated electronic densities of states (DOS) and band structures
are shown in Figs. \ref{dos} and \ref{bands}.
Both BaCu$_2$As$_2$ and SrCu$_2$As$_2$ are metallic.
As may be seen, the Cu $d$ bands are narrow and are located at
high binding energy, more than 3 eV below the Fermi energy, $E_F$.
Furthermore, there is very little Cu $d$ contribution to the DOS
near $E_F$. Therefore, the Cu $d$ orbitals are fully occupied in these
compounds and are chemically inert.
However, it is interesting to note that the Cu bands are somewhat
broader in SrCu$_2$As$_2$ than in BaCu$_2$As$_2$, reflecting the
shorter Cu-Cu distance in the Sr compound (3.02 \AA, vs. 3.14 \AA).
In any case, the Cu $d$ bands lie on
top of a background of broad pnictogen $p$ and metal $sp$ bands extending from
-6 eV to above $E_F$. Therefore these compounds are $sp$ metals.
Turning to the band structure, which is plotted along lines either
in the basal plane (constant $k_z$) or along $k_z$, it is clearly seen
that there are strong dispersions in both the $ab$ plane and along
the $c$-axis ($k_z$) directions, showing three dimensional character.

\begin{figure}[tbp]
\includegraphics[width=0.96\columnwidth,angle=0]{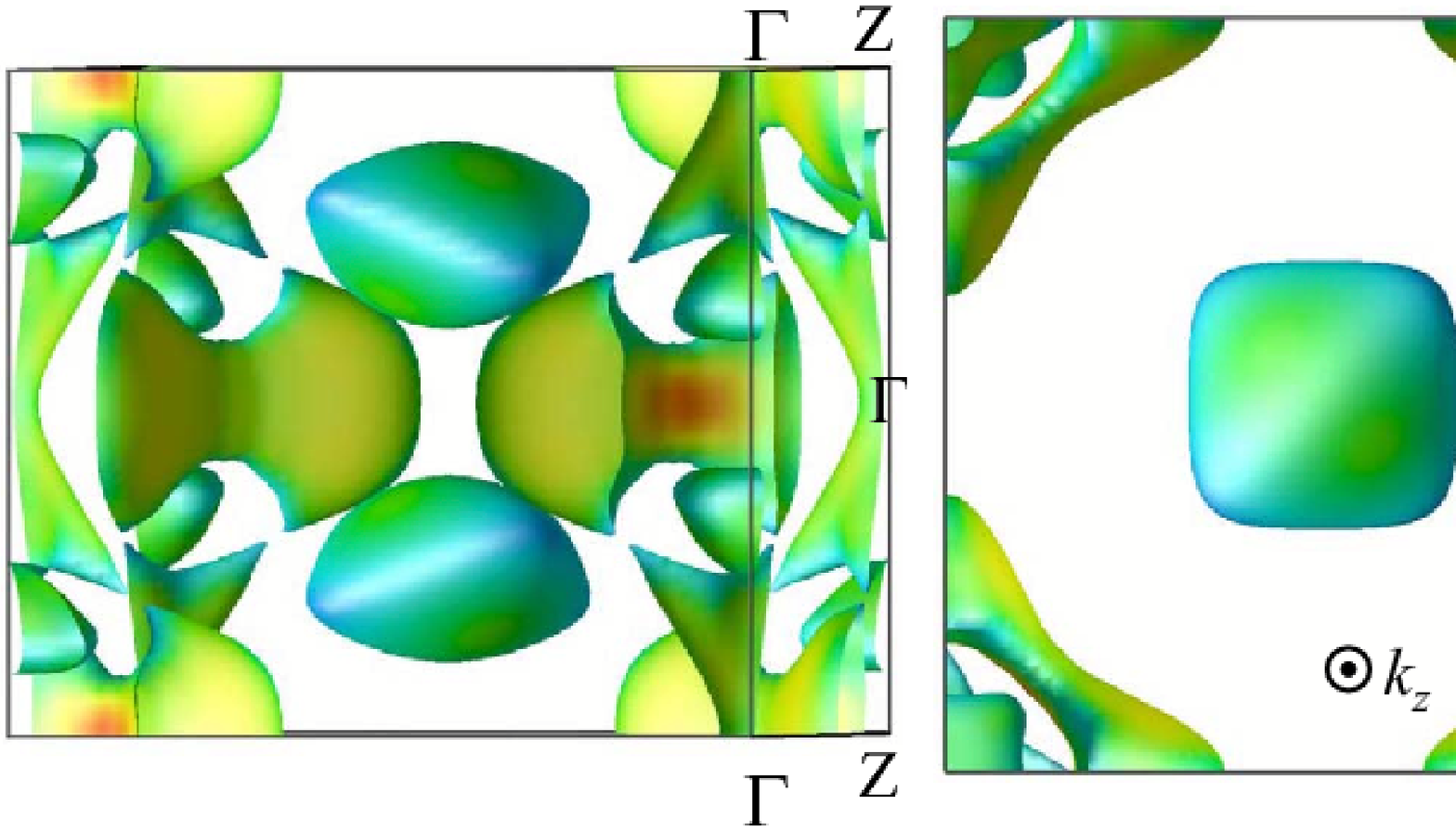}
\caption{(color online)
Fermi surface of BaCu$_2$As$_2$ shaded by velocity (light blue is low
velocity).
The left (right) panel shows a view perpendicular to (along) the
$c$-axis.}
\label{fermi-b}
\end{figure}

\begin{figure}[tbp]
\includegraphics[width=0.96\columnwidth,angle=0]{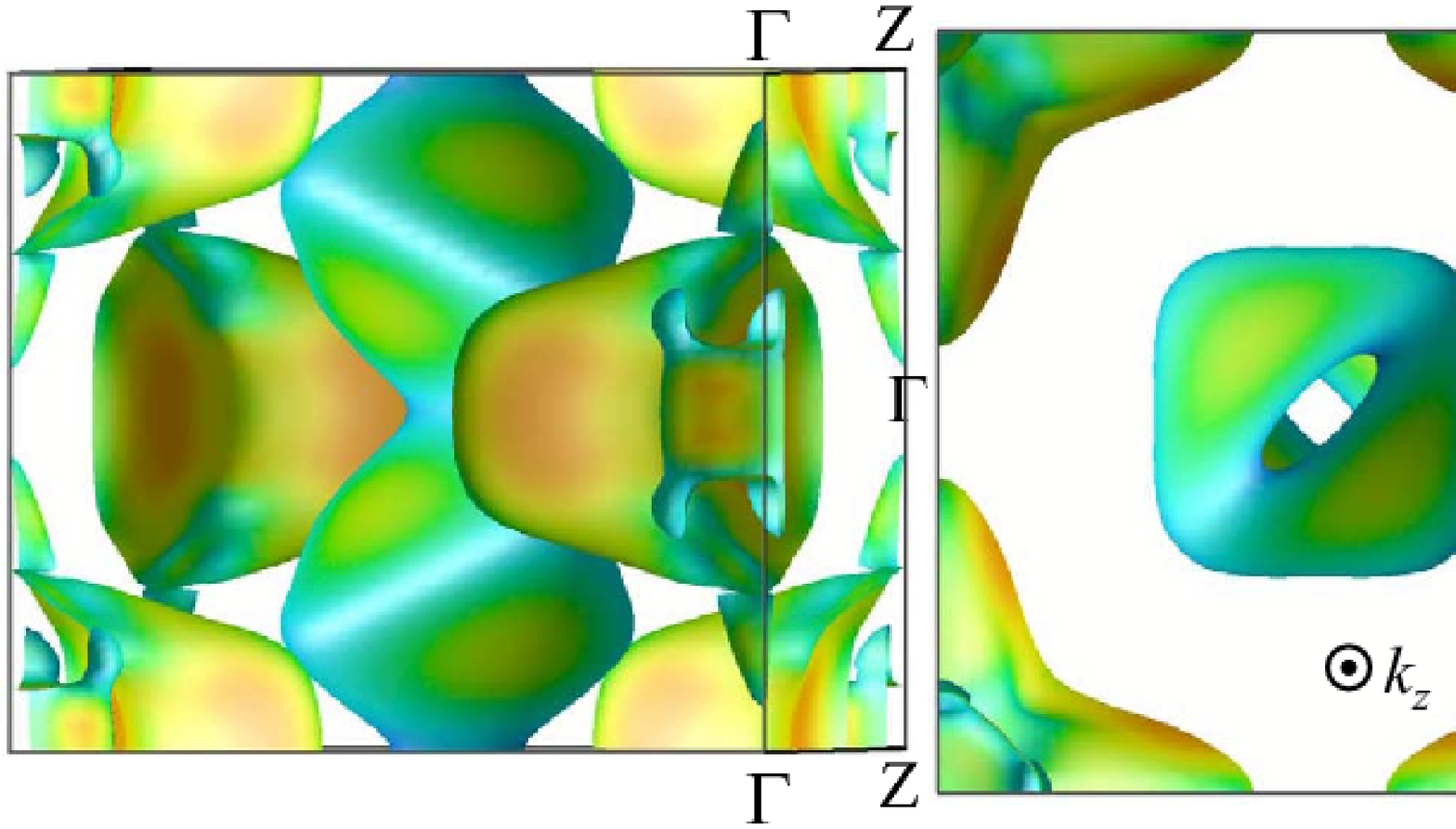}
\caption{(color online)
Fermi surface of SrCu$_2$As$_2$ shaded by velocity (light blue is low
velocity).
The left (right) panel shows a view perpendicular to (along) the
$c$-axis.}
\label{fermi-s}
\end{figure}

The calculated Fermi surfaces for BaCu$_2$As$_2$ and SrCu$_2$As$_2$ are
shown in Figs. \ref{fermi-b} and \ref{fermi-s}, respectively.
The Fermi surfaces are large
and quite three dimensional. The Fermi surface
of BaCu$_2$As$_2$ shows electron pockets along the (1/2,1/2,$k_z$)
direction (the center of the right panel of Fig. \ref{fermi-b}, and 
large complex sheets around (0,0,$k_z$). In SrCu$_2$As$_2$, the
electron pockets are larger and touch forming a large strongly corrugated
electron cylinder running along the zone corner, again accompanied by
complex large Fermi surfaces around the zone center.

\begin{figure}[tbp]
\vspace{0.25cm}
\includegraphics[height=0.96\columnwidth,angle=270]{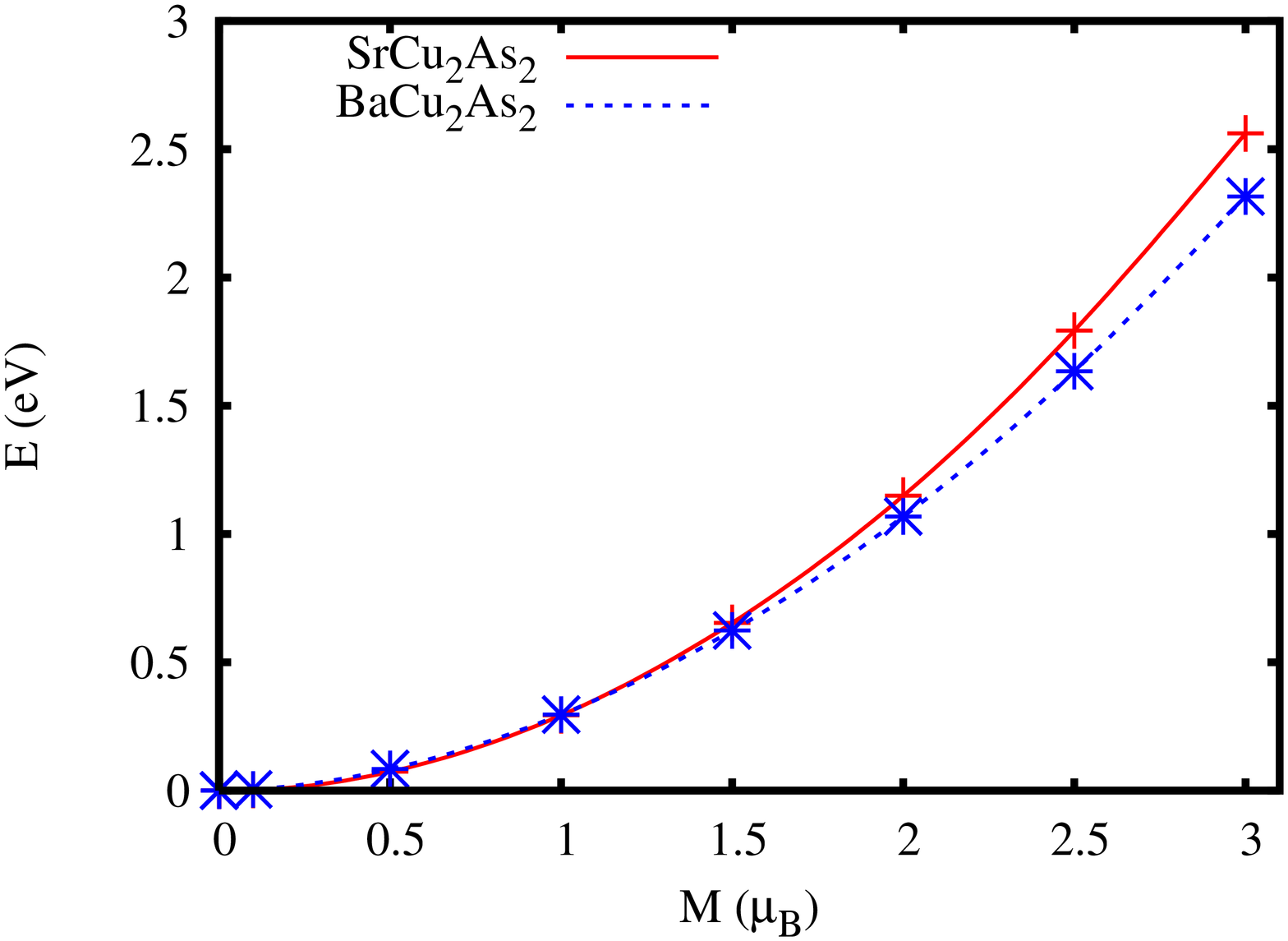}
\caption{(color online)
Fixed spin moment energy as a function of constrained spin magnetization,
shown on a per formula unit basis for BaCu$_2$As$_2$ and SrCu$_2$As$_2$.}
\label{fsm}
\end{figure}

The calculated densities of states at the Fermi energy
are
$N(E_F)$=1.26 eV$^{-1}$ for BaCu$_2$As$_2$ and
$N(E_F)$=1.53 eV$^{-1}$ for SrCu$_2$As$_2$ on a per formula unit basis.
The corresponding Fermi velocities are
$<v_x^2>^{1/2}$=2.8x10$^5$ m/s and $<v_z^2>^{1/2}$=2.6x10$^5$ m/s for
BaCu$_2$As$_2$ and
$<v_x^2>^{1/2}$=2.8x10$^5$ m/s and $<v_z^2>^{1/2}$=2.5x10$^5$ m/s for
SrCu$_2$As$_2$, consistent with weakly anisotropic three dimensional
transport in both compounds.
The high Fermi velocities reflect the dispersive nature of the
$sp$ bands near $E_F$.
The values of $N(E_F)$ are high for an $sp$ metal, but considering
the extended nature of As $p$ orbitals they are far from sufficient to
yield Stoner magnetism. This was confirmed by fixed spin moment calculations
of the energy as a function of constrained spin magnetization, as
shown in Fig. \ref{fsm}. As may be seen, these compounds are not near
magnetism. This is as may be expected considering the absence of
$d$ bands near the Fermi energy.

The general features of the
electronic structures of ThCr$_2$Si$_2$ structure pnictides have
been discussed by Hoffmann and Zheng in terms of an interplay between
pnictogen - pnictogen and pnictogen - metal bonding.
\cite{hoffmann,zheng}
With the additional ingredient of direct metal - metal interactions,
which are important in the middle of the series (Fe, Co, Ni),
\cite{singh-du} this allows a classification of the compounds.
According to our calculations the Cu $d$ states are well below the
Fermi energy, and therefore chemically inert in BaCu$_2$As$_2$ and
SrCu$_2$As$_2$ and the bonding is primarily among the As atoms.
In the Fe, Co and Ni compounds, As is nearly anionic, and metal - metal
bonding is important in addition to weaker pnictogen interactions.
In the Cr and Mn compounds there is strong covalency between the metal $d$
states and the As states, in addition to pnictogen - pnictogen
covalency as evidenced by the three dimensional character of those compounds.
In this regard,
one may note that the $c$-axis lattice parameter of BaCu$_2$As$_2$
is shorter by more than 1.5 \AA, as compared to that of BaNi$_2$As$_2$.
This is highly suggestive that the bonding of these two compounds is
different, and in particular the shorter As - As distances
in the Cu compounds reflects As - As bonding.
This classification also provides an explanation for the fact that
BaFe$_2$Sb$_2$, BaCo$_2$Sb$_2$ and BaNi$_2$Sb$_2$ are not formed.
\cite{pearson,pearson2,just}
In particular, antimony has a greater tendency towards formation of compounds
with pnictogen - pnictogen bonding than arsenic and, as mentioned, in the
Fe, Co and Ni compounds the pnictogens are anionic while metal - metal
bonding is crucial.

To summarize, density functional calculations for BaCu$_2$As$_2$ and
SrCu$_2$As$_2$ show that these materials are $sp$ metals with dispersive
bands at the Fermi energy. The Cu $d$ bands are at high binding energy
in excess of 3 eV, and are therefore inert in these compounds.
These materials are not near magnetism. Therefore, the characteristic
electronic structure associated with the Fe-based superconductors,
in particular an electronic structure associated with anionic pnictogens
and metal - metal $d$ bonding exists in the ThCr$_2$Si$_2$ structure
pnictides for Fe, Co and Ni, but not Cu.
The implication is that while Co and Ni can be used effectively to
dope BaFe$_2$As$_2$ to produce
a coherent alloy with superconductivity,
the other 3$d$ elements, Mn, Cr and Cu will behave
differently. It should be noted
that this last conclusion does not apply to the $\alpha$-PbO
structure chalcogenides, since the bonding and chemistry of chalcogenides
are different than those of arsenides.

\acknowledgments

This work was supported by the Department of Energy,
Division of Materials Sciences and Engineering.

\bibliography{BaCu2As2}
\end{document}